\documentclass[aps,prd,groupedaddress,showpacs,twocolumn]{revtex4}

\usepackage[parfill]{parskip}    
\usepackage{graphicx}
\usepackage{amssymb}
\usepackage{epstopdf}
\usepackage{color}

\usepackage{graphicx}
\usepackage{pdfpages}
\usepackage[colorlinks=true,citecolor=blue,urlcolor=magenta,breaklinks]{hyperref}

\begin{document}

\title{Black hole shadow of a rotating scale--dependent black hole}

\author{Ernesto Contreras 
\footnote{econtreras@usfq.edu.ec}} 
\address{Departamento de F\'isica, Colegio de Ciencias e Ingenier\'ia, Universidad San Francisco de Quito, Quito, Ecuador\\}
\address{School of Physical Sciences \& Nanotechnology, Yachay Tech University, 100119 Urcuqu\'i, Ecuador\\}
\author{\'Angel Rinc\'on {\footnote{angel.rincon@pucv.cl}}}
\address{Instituto de F{\'i}sica, Pontificia Universidad Cat{\'o}lica de Valpara{\'i}so, Avenida Brasil 2950, Casilla 4059, Valpara{\'i}so, Chile\\}

\author{Grigoris Panotopoulos {\footnote{grigorios.panotopoulos@tecnico.ulisboa.pt}}}
\address{Centro de Astrof{\'i}sica e Gravita{\c c}{\~a}o, Departamento de F{\'i}sica, Instituto Superior T{\'e}cnico-IST, Universidade de Lisboa-UL, Av. Rovisco Pais, 1049-001 Lisboa, Portugal\\}

\author{Pedro Bargue\~no {\footnote{pedro.bargueno@ua.es}}}
\address{Departamento de F\'isica Aplicada, Universidad de Alicante, Campus de San Vicente Raspeig, E-030690 Alicante, Spain\\}

\author{Benjamin Koch {\footnote{bkoch@fis.puc.cl}}}
\address{Instituto de F\'{\i}sica,\\ Pontificia Universidad Cat\'olica de Chile\\}

\begin{abstract}
In this work, starting from a spherically symmetric scale--dependent black hole, a rotating solution is obtained by following the Newman--Janis algorithm without complexification. Besides studying the horizon, the static conditions and causality issues of the rotating solution, 
we get and discuss the shape of its shadow. 
\end{abstract}

\maketitle

\section{Introduction}\label{intro}

Black holes (BHs) are a generic prediction of Einstein's General Relativity (GR) and other metric theories of gravity, and they are believed to be formed in the gravitational collapse of massive stars during their final stages. BHs are the simplest objects in the Universe, characterized entirely by a handful of parameters, namely their mass, charges and rotation speed, although they are way more than mathematical objects. Due to Hawking radiation \cite{hawking1,hawking2} and black hole thermodynamics \cite{bekenstein,carter1}, BHs are exciting objects that link together several different disciplines from gravitation and astrophysics to quantum mechanics and statistical physics, and they may provide us with some insight into the quantum nature of gravity.
 
\smallskip

We had to wait 100 years since the formulation of GR \cite{GR} and the first black hole solution obtained by K.~Schwarzschild \cite{SBH} to get the 
strongest evidence so far regarding their existence, and the first image of a black hole. In particular, on one hand, four years ago the LIGO collaboration 
directly detected for the first time the gravitational waves emitted from a BH merger of $\sim 60~M_{\odot}$ \cite{ligo}, but there was no information 
on the defining property of BHs, which is no other than their horizon. On the other hand, just a few weeks ago the Event Horizon Telescope (EHT) 
project \cite{project}, a global very long baseline interferometer array observing at a wavelength of $1.3$~mm, observed a characteristic 
shadow-like image \cite{L1} (see also \cite{L2,L3,L4,L5,L6} for physical origin of the shadow, data processing and calibration, instrumentation etc), 
that is a darker region over a brighter background, via strong gravitational lensing and photon capture at the horizon. Thus, the observation of the 
shadow does probe the spacetime geometry in the vicinity of the horizon, and doing so it tests the existence and properties of the latter \cite{psaltis}, 
although one should keep in mind that other horizonless objects that possess light rings also cast shadows 
\cite{horizonless1,horizonless2,horizonless3,horizonless4,horizonless5,horizonless6,shakih2018,shakih2019},
and therefore the presence of a shadow does not necessarily imply that the object is a BH. Therefore, shadows as well as strong lensing \cite{vib1,vib2} 
images provide us with the exciting opportunity a) to detect the nature of a compact object, and b) to test whether the gravitational field around a 
compact object is described by a rotating or non-rotating geometry. For a recent brief review on shadows see \cite{review}.    

\smallskip

The shadow of the Schwarzschild geometry was considered in \cite{synge,luminet}, while the shadow cast by the Kerr solution \cite{kerr} 
was studied in \cite{bardeen} (see also \cite{monograph}). 
For shadows of Kerr BHs with scalar hair see \cite{carlos1,carlos2}, and for BH shadows in other frameworks see 
\cite{Bambi:2008jg,Bambi:2010hf,study1,study2,Moffat,quint2,study3,study4,study5,study6,bobir2017,kumar2017,kumar2018,ovgun2018,
Konoplya:2019sns,sudipta2019,shakih2019b,contreras2019rot,sabir19,ovgun2019,sunnya,sunnyb}.
Non--rotating solutions have been obtained in non-standard scenarios, such as polytropic BHs \cite{polytropic1,polytropic2} or BHs with quintessential 
energy \cite{quint1}, to mention just a few. 
To explore the physics behind BH shadows, a useful tool is provided by the by now well-known Newman-Janis algorithm (NJA), 
described in \cite{algorithm1,algorithm2}. It is an approach that allows us to generate rotating BH solutions starting from non-rotating seed spacetimes. 
To be more precise, in the present work we will take a variation of the usual NJA, the only difference being the omission of one of the steps of the NJA, namely the complexification of coordinates \cite{quint1}. Instead of this, we will follow an ``alternate" coordinate transformation, which will be explained 
in the next section.

\smallskip

In the present work, we propose to investigate the shadow of a rotating scale--dependent BH. 
The non-rotating, static, spherically symmetric geometry was obtained in \cite{Ben2015}. The metric tensor is a solution to the modified Einstein's field equations in the scale--dependent scenario. In the rotating case, which will be studied in the present work applying the NJA, the computation of the shadow 
would constitute a valuable tool to confirm or refute theoretical predictions regarding the intimate structure of space and time at the strong field regime.

\smallskip

Our work is organized as follows. In the next section, we briefly summarize the main aspects of the NJA, while in section \ref{ng} we study the conditions leading to unstable null trajectories for a generic parametrization of a rotating BH. After that, we review the Scale-Dependent (SD) scenario in the section
\ref{SD}, while in section \ref{rotating} we construct the rotating solution applying the NJA starting from a static, spherically symmetric SD BH, and we investigate some properties of the solution; for instance, horizon and static conditions, causality issues as well as the BH shadow. 
Finally, we conclude our work in the last section. We choose natural units in which $c=1=G$, and we adopt the mostly negative metric signature $(+,-,-,-)$.

\section{Newman--Janis algorithm without complexification}\label{NJ}

The NJA is a solution generating technique which, in its original formulation,
generates rotating metrics starting from static ones. Shortly after its discovery and its use
to re-derive the Kerr metric \cite{algorithm2}, it was successfully used to find the Kerr-Newman spacetime 
\cite{algorithm1}. In particular, the original method uses a complexification of both radial time (and null) coordinates, together with a complex coordinate transformation. At the end of the process,
a change of coordinates to write the result in Boyer--Lindquist coordinates is performed. Although
there are several alternative formulations (apart from the original formulation) for the NJA, 
here we would like to highlight those by Giampieri \cite{Giampieri1990}, Talbot \cite{Talbot1969}, 
Schiffer \cite{Schiffer1973}, Newman \cite{Newman1973},
and Ferraro \cite{Ferraro2014} (see Ref. \cite{Erbin2017} and references therein for a recent and very complete review of the algorithm). 
Regarding the wide belief concerning that the algorithm is a trick without any physical or mathematical basis, we would like to remark the work by 
Drake and Szekeres, who proved that the only perfect fluid generated by the NJA is the (vacuum) Kerr metric and that the only Petrov--D solution to the 
Einstein--Maxwell system is the Kerr--Newman metric \cite{Drake2000}.
More recently, Azreg-A\"{i}nou \cite{azreg2014} has developed a modification of the algorithm to avoid the complexification process, which is usually considered as the main bottleneck of the process due to the absence rigorous statements concerning the possible complexification of the involved functions. Due to this particular feature, in the present manuscript, we will employ this version of the algorithm in the SD context. 
In this section, in order to facilitate the reading of the manuscript, the main aspects of the NJA due to Azreg-A\"{i}nou will be briefly reviewed. 

The starting point is a static spherically symmetric metric parametrized as
\begin{equation}
ds^{2}=G(r)dt^{2}-\frac{dr^{2}}{F(r)}-H(r)d \Omega^2,
\end{equation}
where $d\Omega^2$ is the usual round metric for the 2-sphere.

The next step consists in introducing advanced null coordinates $(u,r,\theta,\phi)$ defined by
\begin{equation}
du=dt-dr/\sqrt{FG},
\end{equation}
from where the non--zero components of the inverse metric can be obtained as
$g^{\mu\nu}=l^{\mu}n^{\nu}+l^{\nu}n^{\mu}-m^{\mu}\bar{m}^{\nu}-m^{\nu}\bar{m}^{\mu}$
with
\begin{eqnarray}
l^{\mu}&=&\delta^{\mu}_{r} ,\\
n^{\mu}&=&\sqrt{F/G}\delta^{\mu}_{u}-(F/2)\delta^{\mu}_{r} ,\\
m^{\mu}&=&(\delta^{\mu}_{\theta}+\frac{i}{\sin\theta}\delta^{\mu}_{\phi})/\sqrt{2H}
\end{eqnarray}
being a set of null tetrads with the following normalization properties: 
$l_{\mu}l^{\mu}=m_{\mu}m^{\mu}=n_{\mu}n^{\mu}=l_{\mu}m^{\mu}=n_{\mu}m^{\mu}=0$ and
$l_{\mu}n^{\mu}=-m_{\mu}\bar{m}^{\mu}=1$. 

After this step, the following complex transformation is introduced:
\begin{eqnarray}
&&r\to r+ia \cos\theta ,\\
&&u\to u-ia \cos\theta.
\end{eqnarray}
Then, if the transformations  
\begin{eqnarray}
&&G(r)\to A(r,\theta, a) ,\\
&&F(r)\to B(r,\theta, a) ,\\
&&H(r)\to \Psi(r,\theta, a) ,
\end{eqnarray}
are considered, we obtain
\begin{eqnarray}
l^{\mu}&=&\delta^{\mu}_{r} ,\\
n^{\mu}&=&\sqrt{B/A}\delta^{\mu}_{u}-(B/2)\delta^{\mu}_{r} ,\\
m^{\mu}&=&(\delta^{\mu}_{\theta}+ia\sin\theta(\delta^{\mu}_{u}-\delta^{\mu}_{r})+\frac{i}{\sin\theta}\delta^{\mu}_{\phi})/\sqrt{2\Psi}.
\end{eqnarray}
Using the previous transformations, the line element can be written in rotating Eddington-–Finkelstein 
coordinates as
\begin{widetext}
\begin{equation} \label{EF}
ds^2 = A du^{2}+2\sqrt{\frac{A}{B}}du dr
  +2a \sin^{2}\theta(\sqrt{\frac{A}{B}}-A)du d\phi \nonumber 
-2a \sin^{2}\theta\sqrt{\frac{A}{B}}dr d\phi-\Psi d\theta^{2} \nonumber
-\sin^{2}\theta(\Psi + a^{2}\sin^{2}\theta(2\sqrt{\frac{A}{B}}-A))d\phi^{2}).
\end{equation}
\end{widetext}
After that, in order to write the metric (\ref{EF}) in Boyer–-Lindquist coordinates, we 
proceed to perform the following global coordinate transformation
\begin{eqnarray}\label{bl}
du &=& dt+\lambda(r)dr ,\\
d\phi &=& d\phi+\chi(r)dr ,
\end{eqnarray}
where $\lambda$ and $\chi$ must depend on $r$ only to ensure the integrability of
Eq. (\ref{bl}). 

The following step lies at the heart of the NJA without complexification \cite{azreg2014}. 
As it is well known, in the original NJA,  the next step in the construction of the rotating metric consists of complexifying the $r$ coordinate. Interestingly, to circumvent it, Azreg-A\"{i}nou proposed \cite{azreg2014} an ansatz for the unknown functions which are involved in the process. Specifically, by taking
\begin{eqnarray}
\lambda &=& -\frac{(K+a^{2})}{FH+a^{2}} ,\\
   \chi &=& -\frac{a}{FH+a^{2}} ,
\end{eqnarray}
where
\begin{equation}
K=\sqrt{\frac{F}{G}}H ,
\end{equation}
and 
\begin{eqnarray}
A(r,\theta)&=&\frac{FH+a^{2}\cos^{2}\theta}{(K+a^{2}\cos^{2}\theta)^{2}}\Psi ,\\
B(r,\theta)&=&\frac{FH+a^{2}\cos^{2}\theta}{\Psi},
\end{eqnarray}
the metric (\ref{EF}) takes a Kerr--like form given by
\begin{widetext}
\begin{equation} \label{blkf}
ds^{2} = \frac{\Psi}{\rho^{2}}\bigg(\frac{\Delta}{\rho^{2}}(dt-a \sin^{2}\theta d\phi)^{2}-\frac{\rho^{2}}{\Delta}dr^{2}-\rho^{2}d\theta^{2}
-\frac{\sin^{2}\theta}{\rho^{2}}(adt-(K+a^{2})d\phi)^{2})\bigg),
\end{equation}
\end{widetext}
with
\begin{equation}
\rho^{2}=K+a^{2}\cos^{2}\theta ,
\end{equation}
where $a=J/M$, with $M,J$ being the mass and the angular momentum, respectively, of the rotating
geometry.

At this point some comments are in order. First, note that although the function $\Psi(r,\theta,a)$ 
remains unknown, it must satisfy the following constraint
\begin{equation}\label{deq}
(K+a^{2}y^{2})^{2}(3\Psi_{,r}\Psi_{,y^{2}}-2\Psi\Psi_{r,y^{2}})=3a^{2}K_{,r}\Psi^{2}
\end{equation}
which corresponds to imposing that the Einstein tensor satisfies $G_{r\theta}=0$. Second, it can be 
shown (see Appendix A of Ref.\cite{azreg2014}) that the metric (\ref{blkf}) satisfies Einstein's
field equations $G_{\mu\nu}=8\pi T_{\mu\nu}$ with a source given by an energy--momentum tensor given by
\begin{equation}\label{tmunu}
T^{\mu\nu}=\varepsilon e^{\mu}_{t}e^{\nu}_{t}+p_{r}e^{\mu}_{r}e^{\nu}_{r}+p_{\theta}e^{\mu}_{\theta}e^{\nu}_{\theta}+p_{\phi}e^{\mu}_{\phi}e^{\nu}_{\phi},
\end{equation}
with 
\begin{eqnarray}
e^{\mu}_{t}&=&\frac{(r^{2}+a^{2},0,0,a)}{\sqrt{\rho^{2}\Delta}}\label{tet1} ,\\
e^{\mu}_{r}&=&\frac{\sqrt{\Delta}(0,1,0,0)}{\sqrt{\rho^{2}}}\label{tet2} ,\\
e^{\mu}_{\theta}&=&\frac{(0,0,1,0)}{\sqrt{\rho^{2}}}\label{tet3} ,\\
e^{\mu}_{\phi}&=&-\frac{(a\sin^{2}\theta,0,0,1)}{\sqrt{\rho^{2}}\sin\theta}\label{tet4}.
\end{eqnarray}
Finally, in order to ensure the consistency of Einstein's field equations, the unknown 
$\Psi$ must satisfy an extra constraint given by
\begin{widetext}
\begin{equation}\label{econs}
(K_{,r}^{2}+K(2-K_{,rr})-a^{2}y^{2}(2+K_{,rr}))\Psi 
+(K+a^{2}y^{})(4y^{2}\Psi_{y^{2}}-K_{,r}\Psi_{,r})=0 .
\end{equation}
\end{widetext}
It is worth mentioning that, in terms of a general function,  $\Psi_{g}$, that solves Eqs. (\ref{deq}) and (\ref{econs}), namely
\begin{eqnarray}
\Psi_{g}=He^{a^{2}\psi},
\end{eqnarray}
with $\psi\equiv\psi(r,y^{2},a^{2})$,	
the components of the energy--momentum tensor can be written as (see, for example, Ref. \cite{arz2014})
\begin{widetext}
\begin{eqnarray}
\varepsilon&=&\frac{1}{\Psi_{g}}
-\frac{a^{2}(20y^{2}(K+a^{2})+24 y^{2}f+(1-y^{2})K_{,r}^{2})}{4\Psi_{g}\rho^{4}}+\frac{3\Delta(H_{,r}+a^{2}H\psi_{,r})^{2}-4a^{4}y^{2}(1-y^{2})H^{2}\psi_{,y^{2}}^{2})}{4H^{2}\Psi_{g}}\nonumber\\
&&+\frac{2 a^{2}(a^{2}y^{2}(1+y^{2})-(1-3y^{2})K)\psi_{,y^{2}}}{\Psi_{g}\rho^{2}}-\frac{1}{2H\Psi_{g}}(
8a^{2}y^{2}(1-y^{2})H\Psi_{,y^{2}y^{2}}+\Delta_{,r}(H_{,r}+a^{2}H\psi_{,r})\nonumber\\
&&+2\Delta(H_{,rr}+a^{2}(2H_{,r}\psi_{,r}+H(a^{2}\psi_{,r}^{2}+\psi_{,rr}))))+\frac{2a^{2}}{\Psi_{g}\rho^{2}}\label{den}\\
p_{r}&=&-\varepsilon+\frac{2a^{2}y^{2}\Delta}
{\Psi_{g}\rho^{4}}-\frac{\Delta(H_{,r}K_{,r}+a^{2}HK_{,r}\psi)}{H\Psi_{g}\rho^{2}}+
\frac{\Delta}{2H^{2}\Psi_{g}}(3H_{,r}^{2}-2HH_{,rr}\psi_{r}+2a^{2}HH_{,r}\psi_{,r}+a^{4}H^{2}\psi_{,r}^{2}) \nonumber \\
&-&2a^{2}H^{2}\psi_{,rr}\label{pr}\\
p_{\theta}-p_{\phi}&=&\frac{a^{2}(1-y^{2})K_{,r}^{2}}{2\Psi_{g}\rho^{4}}-\frac{4a^{4}(1-y^{2})y^{2}\psi_{,y^{2}}}{\Psi_{g}\rho^{2}}+\frac{2a^{2}(1-y^{2})(a^{2}y^{2}\psi_{,y^{2}}^{2}-2y^{2}\psi_{,y^{2}y^{2}}-\psi_{,y^{2}})}{\Psi_{g}}\label{pp}.
\end{eqnarray}
\end{widetext}
with
\begin{eqnarray}\label{defin}
\rho^{2}&=&r^{2}+a^{2}\cos^{2}\theta ,\\
2f & = & r^{2}(1-F) ,\\
\Delta & = &-2f+a^{2}+ r^2 , \\
\Sigma & = & (r^{2}+a^{2})^{2}-a^{2} \Delta \sin^{2} \theta.
\end{eqnarray}
In this sense, the Einstein field equations are satisfied whenever the matter sector is given by the set of equations 
(\ref{den}), (\ref{pr}), (\ref{pp}).

After implementing the protocole here reviewed, the above expressions can be simplified in the 
particular case $G=F$ and $H=r^{2}$. In this case it can be shown that one possible solution of 
Eq. (\ref{deq}) is given by
\begin{equation}
\Psi=r^{2}+a^{2}\cos^{2}\theta
\end{equation}
and, then, the metric given by (\ref{blkf}) takes the form
\begin{widetext}
\begin{equation}\label{blsm}
ds^{2}=\left(1-\frac{2f}{\rho^{2}}\right)dt^{2}-\frac{\rho^{2}}{\Delta}dr^{2}
+\frac{4af \sin^{2}\theta}{\rho^{2}}dtd\phi\nonumber\\
-\rho^{2}d\theta^{2}-\frac{\Sigma \sin^{2}\theta}{\rho^{2}}d\phi^{2} ,
\end{equation}
\end{widetext}
showing that the non--rotating black hole solution is recovered when $a=0$. Note also that, in this particular case, Eqs. (\ref{den}), (\ref{pr}) and (\ref{pp})
 reduce to (see, Ref. \cite{arzplb})
\begin{eqnarray}
\varepsilon&=&\frac{2(rf'-f)}{\rho^{4}}\label{ep}\\
p_{r}&=&-\varepsilon\label{pre}\\
p_{\theta}&=&-p_{r}-\frac{f''}{\rho^{2}}\label{pthe}\\
p_{\phi}&=&p_{\theta}\label{pfi}
\end{eqnarray}

Before concluding this section we would like to highlight a couple of points. First, the NJ algorithm summarized here (originally developed in \cite{azreg2014} as stated before) does not appear to present pathologies \cite{rodrigues2017,bobirmod}. However, as claimed by Rodrigues and Junior in \cite{rodrigues2017} some inconsistencies could appear depending on the way the method is used to generate rotating solutions. For example, if the action contains a matter Lagrangian, the correct way to implement the method could be to integrate the equations of motion associated with variations of the action respect to the matter field and then to insert this result into the Einstein equations projected with tetrads given by Eqs. (\ref{tmunu}), (\ref{tet1}), (\ref{tet2}), (\ref{tet3}) and (\ref{tet4}). On the contrary, the final result must be considered as an approximation of the real rotating solution. We shall discuss more about this point in the context of scale--dependent gravity at the end of section \ref{rotating}.  Second, 
the algorithm is not restricted to asymptotically flat static seeds. Indeed,
the only requirement to obtain rotating solutions using the algorithm consists in imposing spherical symmetry for the corresponding 
static solution. As an example for this assertion, please see Ref. \cite{arzplb}, where the steps here developed are used to 
obtain an asymptotically de Sitter rotating solution.

\section{Null geodesics around the rotating black hole}\label{ng}
In this section, we use the well known Hamilton--Jacobi formalism to obtain the null geodesic equations in a rotating space--time to find the celestial coordinates to study the shadow of a generic rotating metric.
From the Hamilton--Jacobi equations \cite{carter}
\begin{equation}\label{jac}
\frac{\partial S}{\partial\tau}=\frac{1}{2}g^{\mu\nu}\partial_{\mu}S\partial_{\nu}S ,
\end{equation}
where $\tau$ is the proper time and $S$ is the Jacobi action and assuming that Eq. (\ref{jac})
is separable, Eq. (\ref{jac}) reads
\begin{equation}\label{separable}
S=-Et+\Phi\phi+S_{r}(r)+S_{\theta}(\theta) ,
\end{equation}
with $E$ and $\Phi$ being the conserved energy and angular momentum respectively. Plugging
(\ref{separable}) in (\ref{jac}) we verify that
\begin{eqnarray}
S_{r} & = & \int\limits^{r}\frac{\sqrt{R(r)}}{\Delta}dr ,\\
S_{\theta} & = & \int\limits^{\theta}\sqrt{\Theta(\theta)}d\theta,
\end{eqnarray}
where
\begin{eqnarray}
R &=& ((r^{2}+a^{2})E-a\Phi)^{2}-\Delta(Q+(\Phi-aE)^{2}),\hspace{10pt} \\
\Theta(\theta)&=& Q-(\Phi^{2}\csc^{2}\theta-a^{2}E^{2})\cos^{2}\theta ,
\end{eqnarray}
where $Q$  is the Carter constant. Now, to obtain the unstable photon orbits in the rotating
space--time we impose $R=0$ and $R'=0$, namely
\begin{eqnarray}
\left(a^2-a \xi +r^2\right)^2-\left(a^2+r^2 F\right) \left((a-\xi )^2+\eta \right) = 0,&&  \\
4 \left(a^2-a \xi +r^2\right)-\left((a-\xi )^2+\eta \right) \left(r F'+2 F\right) = 0,&&
\end{eqnarray}
where $\xi=\Phi/E$ and $\eta=Q/E^{2}$ are the impact parameters. Accordingly, 
\begin{eqnarray}
\xi &=& -\frac{4 \left(a^2+r^2 F\right)}{a(r F'+2 F)}+a+\frac{r^2}{a}\label{chi}, \\
\eta &=& \frac{r^3 \left(8 a^2 F'-r \left(r F'-2 F\right)^2\right)}{a^2 \left(r F'+2 F\right)^2}\label{eta}.
\end{eqnarray}
In the above expressions, $r$ stands for the radius of the unstable null orbits.

Finally, the apparent shape of the shadow can be obtained from the celestial coordinates which are given by \cite{vazquez}
\begin{eqnarray}
\alpha&=&\lim\limits_{r_{0}\to \infty}\left(-r_{0}^{2}\sin\theta_{0}\frac{d\phi}{dr}\bigg|_{(r_{0,\theta_{0}})}\right)
\label{alpha} , \\
\beta&=&\lim\limits_{r_{0}\to\infty}\left(r_{0}^{2}\frac{d\theta}{dr}\bigg|_{(r_{0},\theta_{0})}\right)
\label{beta},
\end{eqnarray}
where $(r_{0},\theta_{0})$ are the coordinates of the observer.  As we shall dicuss in the next section, our seed solution is aymptocally anti de Sitter. In this sense, as stated in \cite{grezenbach} in the context of shadows of Kerr–Newman–NUT black holes with a cosmological constant, the case $\Lambda\le0$
the domain of outer communication of the black hole, namely, the region between $r =\infty$ and the first horizon corresponds to the region where we can place observers for observing the shadow of the black hole.

\section{Scale--dependence in black hole physics}
\label{SD}

This section is devoted to reviewing the essential ingredients of the now well--known SD gravity. The motivation of this alternative approach in which the couplings evolve with respect to an arbitrary scale, which is best understood in the context of quantum gravity \cite{Rovelli:2007uwt}. 
Up to now a ``consistent and predictive" description of quantum gravity is doubtlessly an open task. 
Among the numerous attempts to theoretically tackle the problem of quantum gravity, the Asymptotic Safety scenario \cite{Weinberg:1976xy} which is using the tool of effective quantum actions and the technique of Exact Renormalization Groups (ERG), has generated promising results and increasing interest~\cite{Wetterich:1992yh,Morris:1993qb,Reuter:1996cp,Reuter:2001ag,Litim:2002xm,Litim:2003vp,Niedermaier:2006wt,Niedermaier:2006ns,Gies:2006wv,Machado:2007ea,Percacci:2007sz,Codello:2008vh,Benedetti:2009rx,Manrique:2009uh,Manrique:2010am,Manrique:2010mq,Eichhorn:2010tb,Litim:2011cp,Falls:2013bv,Dona:2013qba,Falls:2014tra,Eichhorn:2018yfc,Eichhorn:2017egq}. 
In those effective actions, the quantum features appear through the different field operators and most importantly, the running of the corresponding coupling constants.  
The scale--dependent effective average action, which is the final result of many quantum approaches is the starting point of the SD approach.
The idea of the SD approach is to obtain self-consistent quantum background solutions of the gap equations, derived from an effective average action. This approach is complementary to the improving solutions approach, where solutions of the classical field equations are improvement
by promoting coupling constants to running coupling constants~\cite{Bonanno:1998ye,Bonanno:2000ep,Emoto:2005te,Bonanno:2006eu,Reuter:2006rg,Koch:2007yt,Hewett:2007st,Litim:2007iu,Burschil:2009va,Falls:2010he,Casadio:2010fw,Reuter:2010xb,Cai:2010zh,Falls:2012nd,Becker:2012js,Koch:2013owa,Koch:2014cqa,Gonzalez:2015upa,Bonanno:2016dyv,Torres:2017ygl,Bonanno:2017zen,Pawlowski:2018swz,Platania:2019kyx,Held:2019xde}.

Even though SD gravity is quite recent (see \cite{Koch:2010nn,Contreras:2013hua,Koch:2014joa,Koch:2015nva,Rodrigues:2015rya,Koch:2016uso,Rincon:2017ypd,Rincon:2017goj,Rincon:2017ayr,Contreras:2017eza,Rincon:2018sgd,Hernandez-Arboleda:2018qdo,Contreras:2018dhs,Rincon:2018dsq,Contreras:2018gct,Contreras:2018gpl,Contreras:2018swc,
Canales:2018tbn,Rincon:2019cix,Rincon:2019zxk,Fathi:2019jid,Contreras:2019fwu}), approaches where at least one of the parameters evolves indeed exist in the literature. 
One of the most famous of them is the Brans--Dicke (BD) theory~\cite{Kang:1996rj}. In such theory, the Newton coupling $G_0$ is taken to be an arbitrary scalar field $\phi$ via the simple replacement $G \rightarrow \phi^{-1}$. Although the BD theory is a classical theory, the computation is similar to the SD approach but the foundation if quite different.

The solutions obtained from the SD scenario are generalizations of the classical solutions,
which can be recovered by setting the parameter controlling the amount of scale dependence to zero.
In the simplest case, one only has two couplings: i) Newton's coupling $G_k$ and ii) the cosmological coupling $\Lambda_k$. As usual, we can define the auxiliary parameter $\kappa_k \equiv 8 \pi G_{k}$. 
The ``dynamical'' fields are the metric field $g_{\mu \nu}$ and the arbitrary renormalization scale $k$.

The effective action is then written as
	\begin{equation}\label{actionG}
	\Gamma[g_{\mu \nu}, k] \equiv \int \mathrm{d}^4 x \sqrt{-g}
	\Bigg[ 
	\frac{1}{2 \kappa_k} \Bigl(R - 2 \Lambda_k \Bigl) \ + \ \mathcal{L}_M
	\Bigg]   ,
	\end{equation}

where $\mathcal{L}_M$ is the Lagrangian density of the matter fields. The variation of the effective average action with respect to the metric field gives the effective Einstein field equations
	\begin{equation}\label{gmn}
	G_{\mu \nu } + \Lambda_k g_{\mu \nu} \equiv \kappa_k T_{\mu \nu}^{\text{effec}},
	\end{equation}
and the effective energy--momentum tensor satisfying
	\begin{equation}
	\kappa_k T_{\mu \nu}^{\text{effec}} =  \kappa_k T_{\mu \nu}^{M} - \Delta t_{\mu \nu}.
	\end{equation}
It is mandatory to point out that the effective energy--momentum tensor takes into account two contributions: i) the usual matter content and ii) the non--matter source (provided by the running of the gravitational coupling).
The additional tensor is 
	\begin{equation}\label{Gg}
	\Delta t_{\mu \nu} \equiv G_k \Bigl( g_{\mu \nu} \square - \nabla_{\mu} \nabla_{\nu} \Bigl) G_k^{-1}. 
	\end{equation}
Given that the purpose of this paper is to investigate a SD BH in four--dimensional spacetime using the NJA formalism, we set  $T_{\mu \nu}^{M} = 0$ (which implies $\mathcal{L}_{M}=0$ in the action (\ref{actionG})) for simplicity, although a matter source is always an interesting ingredient in gravitational theories.

Now, varying the effective average action respect the additional field $k(x)$, we obtain an auxiliary equation to complete the set. The last condition reads:
	\begin{equation}\label{eomk}
	    \frac{\delta \Gamma[g_{\mu \nu},k]}{\delta k} = 0.
	\end{equation}
	
	This restriction can be seen as an a posteriori condition towards background independence 
\cite{Stevenson:1981vj,Reuter:2003ca,Becker:2014qya,Dietz:2015owa,Labus:2016lkh,Morris:2016spn,Ohta:2017dsq}.
    The condition (\ref{eomk}) gives an immediate link
 between $G_k$ and $\Lambda_k$. In such, we observe that the cosmological parameter is mandatory to obtain self--consistent SD solutions.  As we commented before, the above equation closes the system.
    
Now, a crucial point in many approaches to SD 
 arises when one circumvents a particular choice of $k(x)$. 
This is a reasonable thing to do
because, in general, when one naively identifies the renormalization scale $k(x)$ in terms of the physical variables of the system under consideration 
$k\rightarrow k(x, \dots)$, the reparametrization symmetry is not preserved anymore. 
One can then complete the set of equations of motion
by assuming some energy constraints. 
It is very well--known that the energy conditions can be violated, but, in general, a physical solution maintains the validity of, at least, one of the energy conditions. 
In our case, we will use the null energy condition (NEC) because it is the least restrictive of them. In the extreme case, we have
	\begin{equation}\label{eomcond}
	    T_{\mu \nu}^{\text{effec}} \ell^\mu \ell^\nu  = -\Delta t_{\mu\nu}\ell^\mu \ell^\nu =^{!} 0,
	\end{equation}
	where $\ell^{\mu}$ is a null vector, similar to that considered in \cite{Rincon:2017ayr}. 
	For the line element
\begin{equation}\label{metric}
ds^{2}=\mathcal{F}dt^{2}-\mathcal{F}^{-1} d r^{2}-r^{2}d \Omega^{2},
\end{equation}
	A clever choice of this vector let us to obtain a differential equation for the gravitational coupling, namely
	\begin{equation} \label{EDO_G}
	    G(r)\frac{\mathrm{d}^2G(r)}{\mathrm{d}r^2} - 2 \left(\frac{\mathrm{d}G(r)}{\mathrm{d}r}\right)^2 = 0,
	\end{equation}
the solution of the above differential equation is given by
\begin{eqnarray}\label{G}
G(r)=\frac{G_{0}}{1+\epsilon r},
\end{eqnarray}
and serves to decrease the number of unknown functions of the problem. After replacing $G(r)$ into the effective Einstein field equations we then obtain the functions $\Lambda(r)$ and $\mathcal{F}(r)$~\cite{Ben2015}.
One finds \footnote{Indeed, in Ref. \cite{Ben2015}, the logarithmic term is $\log \left(\frac{c_{4}(\epsilon r  +1)}{ r}\right)$ with
$c_{4}$ an arbitrary constant which value is discussed in  \cite{Ben2015}. However, a dimensional analysis reveals that
$c_{4}$ can be associated to the running parameter $\epsilon$.}
\begin{widetext}
\begin{eqnarray}\label{sd}
\mathcal{F}=1-\frac{2 G_{0} M}{r}-\frac{\Lambda  r^2}{3}
-r \epsilon  (6 G_{0} M \epsilon +1)+3 G_{0} M \epsilon    
+ r^2 \epsilon ^2 (6 G_{0} M \epsilon +1) \log \left(1+\frac{1}{\epsilon r}\right),
\end{eqnarray}
\end{widetext}
and
\begin{widetext}
\begin{eqnarray}\label{L}
\Lambda(r) &=& \frac{6 G_0 M_0 \epsilon ^2-72 r \epsilon ^2 \left(r \epsilon +\frac{1}{2}\right) (r \epsilon +1) \left(G_0 M_0 \epsilon +\frac{1}{6}\right) \log \left(\frac{r \epsilon +1}{r \epsilon }\right) 
+
r \left(72 G_0 M_0 \epsilon ^3+2 \Lambda _0+11 \epsilon ^2\right)
}{2 r (r \epsilon +1)^2}
\nonumber
\\
&+&
\frac{4 \Lambda _0 r^3 \epsilon ^2 + r^2 \left(72 G_0 M_0 \epsilon ^2+12 \epsilon ^3+6 \Lambda _0 \epsilon \right)}{2 r (r \epsilon +1)^2}
\end{eqnarray}
\end{widetext}
It is worth noticing that, one of the integration constants, that appears when solving (\ref{eomcond})
can be identified as the running parameter $\epsilon$, which measures the strength of the scale--dependence of the modified solution. 
Even more, we expect that the improved SD solution deviates slowly from the classical one so that $\epsilon$ corresponds to a small quantity.  Indeed, in the limit $\epsilon\to 0$ the classical solution is recovered, namely $G(r)\to G_{0}$, with $G_{0}$ the Newton constant
and $\mathcal{F}(r)\to \mathcal{F}_0(r)$ as in the classical SD Schwarzschild--(anti) de Sitter space--time.\\
   It is remarkable that for the case of coordinate transformations we have
	\begin{equation}
	\nabla^\mu G_{\mu \nu}=0.
	\end{equation}	
By inserting the solution, into $\Gamma_k$ and applying the chain rule
for $k=k(r)$ one can readily confirm that variational implementation of scale independence (\ref{eomk}) is indeed fulfilled by the solution.

A surprising feature of the above solution is the behavior for a very large and very small radius. This asymptotic behavior reflects the gravitational instability predicted and found in certain AS scenarios~\cite{Wetterich:2017ixo,Biemans:2016rvp,Biemans:2017zca,Rincon:2019cix}
and not the  $k\sim 1/r$ relation one would obtain from a naive dimensional guess.


\section{Rotating scale--dependent black hole solution}\label{rotating}

In this section we use the NJ algorithm to construct the rotating BH solution from a static and spherically symmetric 
SD BH. Then  the properties of this rotating SD space-time are studied.
Since we are mostly interested in the black hole shadow
we will leave the transcription of $G(r)$ and $\Lambda(r)$ to the corresponding
$G(r,\theta)$ and $\Lambda(r,\theta)$ to a future study.

As a starting point, we consider the SD BH solution (\ref{sd}) obtained in \cite{Ben2015}. Note that, using this metric function, the rotating metric reads as Eq. (\ref{blsm}) where we must replace $F$ by $\mathcal{F}$ in all the definitions. Moreover, from Eqs. (\ref{ep}), (\ref{pre}), (\ref{pthe}) and (\ref{pfi}), the matter sector which solve the Einstein field equations reads
\begin{widetext}
\begin{eqnarray}
\varepsilon&=&\frac{r^2 \left(-3 r^2 \epsilon ^2 (r \epsilon +1) (6 G_{0} M \epsilon +1) \log \left(\frac{1}{r \epsilon }+1\right)+3 G_{0} M \epsilon  (3 r \epsilon  (2 r \epsilon +1)-1)+r (\Lambda  r (r \epsilon +1)+\epsilon  (3 r \epsilon +2))\right)}{(r \epsilon +1) \left(a^2 \cos ^2(\theta )+r^2\right)^2}\\
p_{\theta}&=&\frac{1}{2 (r \epsilon +1)^2 \left(a^2 \cos ^2(\theta )+r^2\right)^2}\bigg(
6 r^2 \epsilon ^2 (r \epsilon +1)^2 (6 \text{G0} M \epsilon +1) \log \left(\frac{1}{r \epsilon }+1\right) \left(2 a^2 \cos ^2(\theta )+r^2\right)\nonumber\\
&&-a^2 \cos ^2(\theta ) \left(6 \text{G0} M \epsilon  (2 r \epsilon +1) (6 r \epsilon  (r \epsilon +1)-1)+r \left(4 \Lambda  r (r \epsilon +1)^2+\epsilon  (r \epsilon  (12 r \epsilon +19)+6)\right)\right)\nonumber\\
&&+r^3 \left(-\epsilon  (3 r \epsilon  (2 r \epsilon +3)+2) (6 \text{G0} M \epsilon +1)-2 \Lambda  r (r \epsilon +1)^2\right)
\bigg)
\end{eqnarray}
\end{widetext}
It is noticeable that in the limit $\epsilon\to 0$ the matter sector of the rotating solution correponds to the de Sitter solution found in \cite{arzplb}, namely,
\begin{eqnarray}
\varepsilon&=&\frac{\Lambda  r^4}{\left(a^2 \cos ^2(\theta )+r^2\right)^2}\\
p_{\theta}&=&-\frac{\Lambda  r^2 \left(2 a^2 \cos ^2(\theta )+r^2\right)}{\left(a^2 \cos ^2(\theta )+r^2\right)^2}.
\end{eqnarray}

\begin{figure}[h!]
\centering
\includegraphics[scale=0.5]{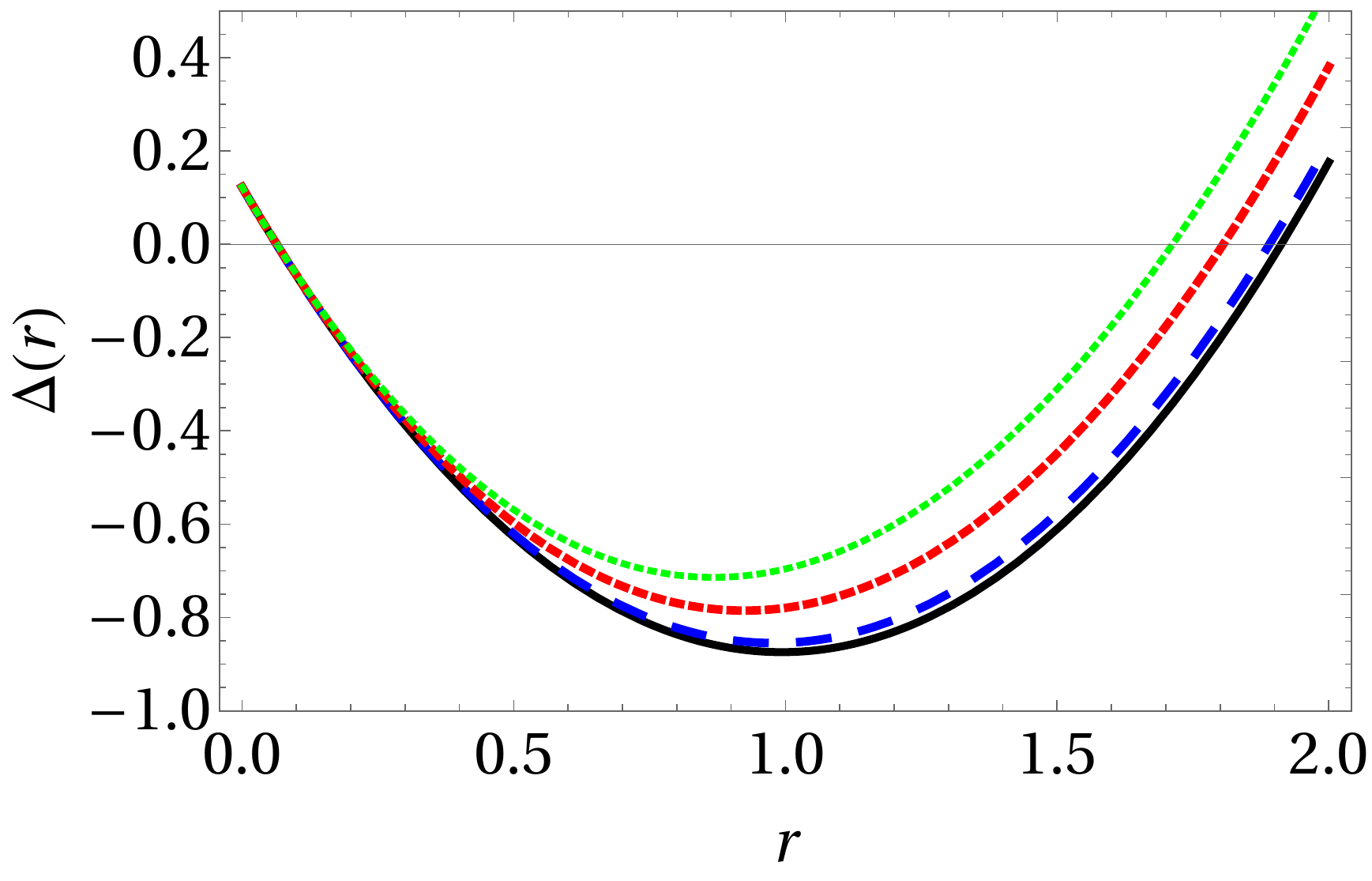}
\caption{\label{fig1} 
Behavior of $\Delta(r)$ for 
 $a=0.35$, $\Lambda=-0.01$, $M=1$ and $G_{0}=1$
with $\epsilon=0.00$ (black solid line), $\epsilon=0.01$ (blue dashed line),
$\epsilon=0.05$ (red dotted line) and $\epsilon=0.10$ (green dotted line)
}
\end{figure}

\begin{figure}[h!]
\centering
\includegraphics[scale=0.45]{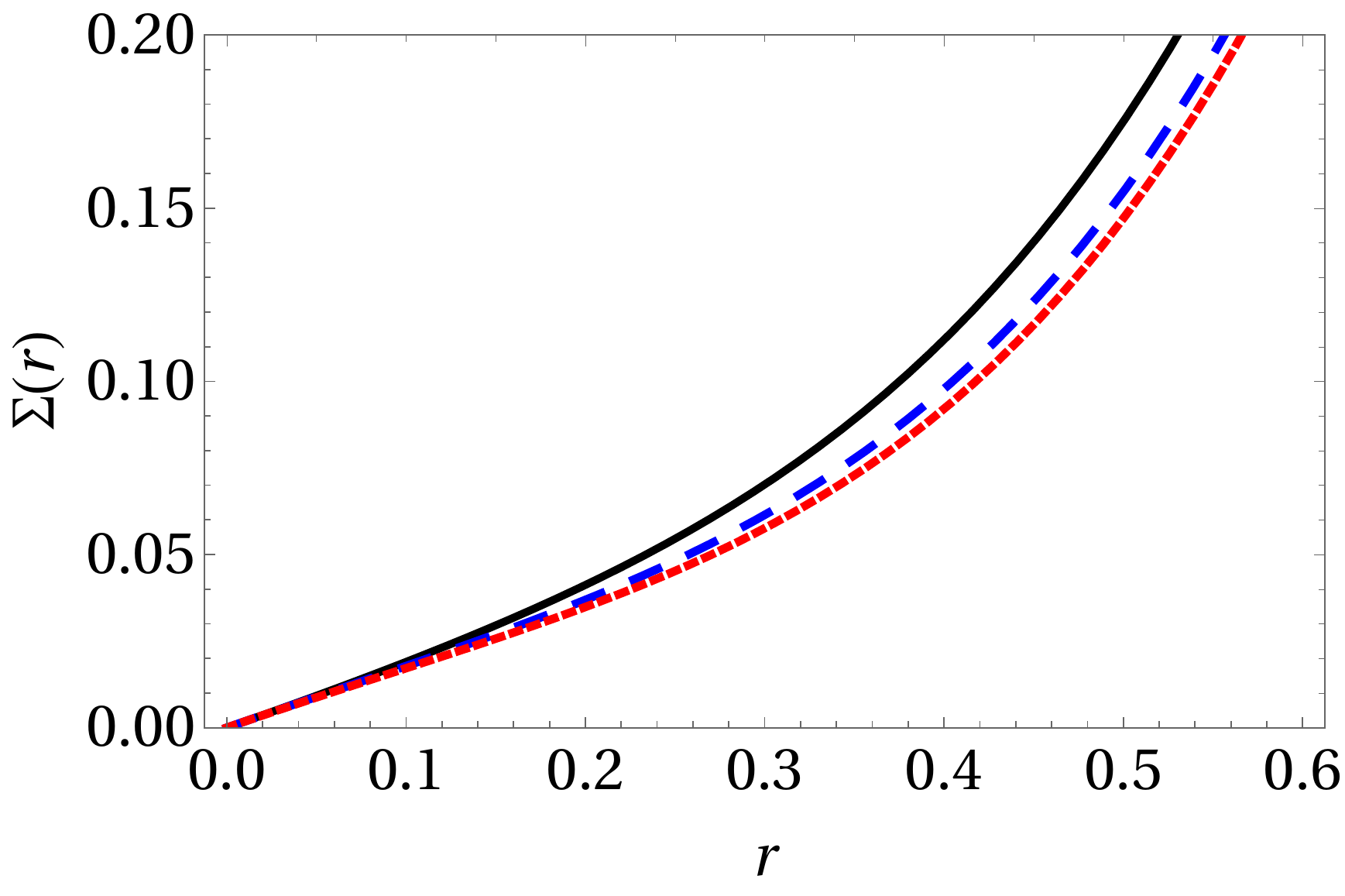}
\caption{\label{figvc} 
Behaviour of $\Sigma$ by the rotating SD BH for $a=0.35$, $\Lambda=-0.01$, $M=1$ and $G_{0}=1$ with $\epsilon=0.00$ (black solid line), $\epsilon=0.50$ (blue dashed line) and
$\epsilon=0.80$ (red dotted line)
 }
\end{figure}

\begin{figure*}[hbt!]
\centering
\includegraphics[scale=0.5]{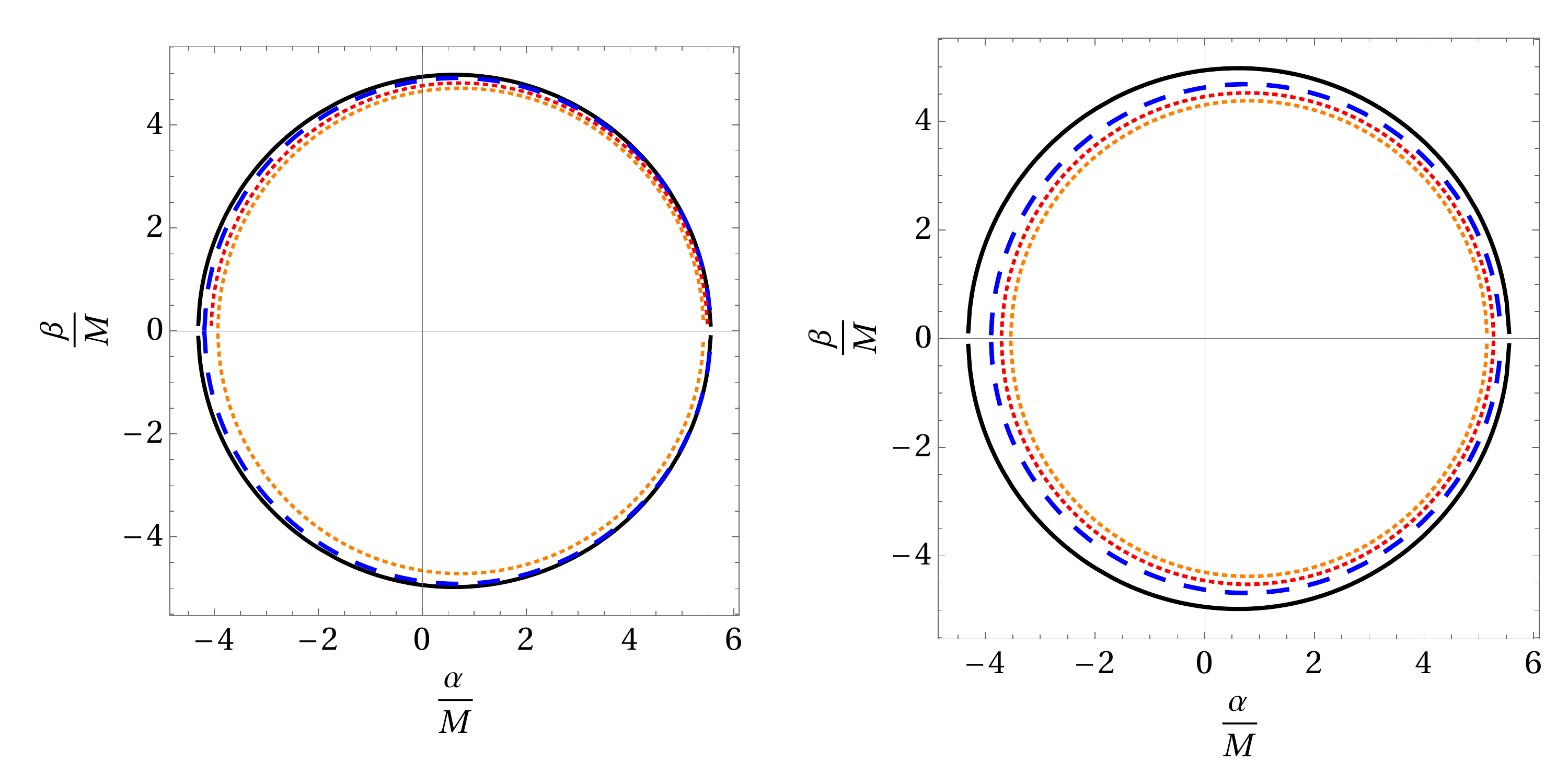}
\caption{\label{fig2} 
Silhouette of the shadow cast by the rotating SD BH for $a=0.35$ with $\Lambda=-0.01$, $M=1$ and $G_{0}=1$. In the left panel $\epsilon=0.00$ (black solid line), $\epsilon=0.03$ (blue dashed line) and
$\epsilon=0.06$ (red dotted line) and $\epsilon=0.09$ (orange dotted line). In the right panel
$\epsilon=0.00$ (black solid line), $\epsilon=0.10$ (blue dashed line) and
$\epsilon=0.15$ (red dotted line) and $\epsilon=0.20$ (orange dotted line)
 }
\end{figure*}

\begin{figure*}[hbt!]
\centering
\includegraphics[scale=0.5]{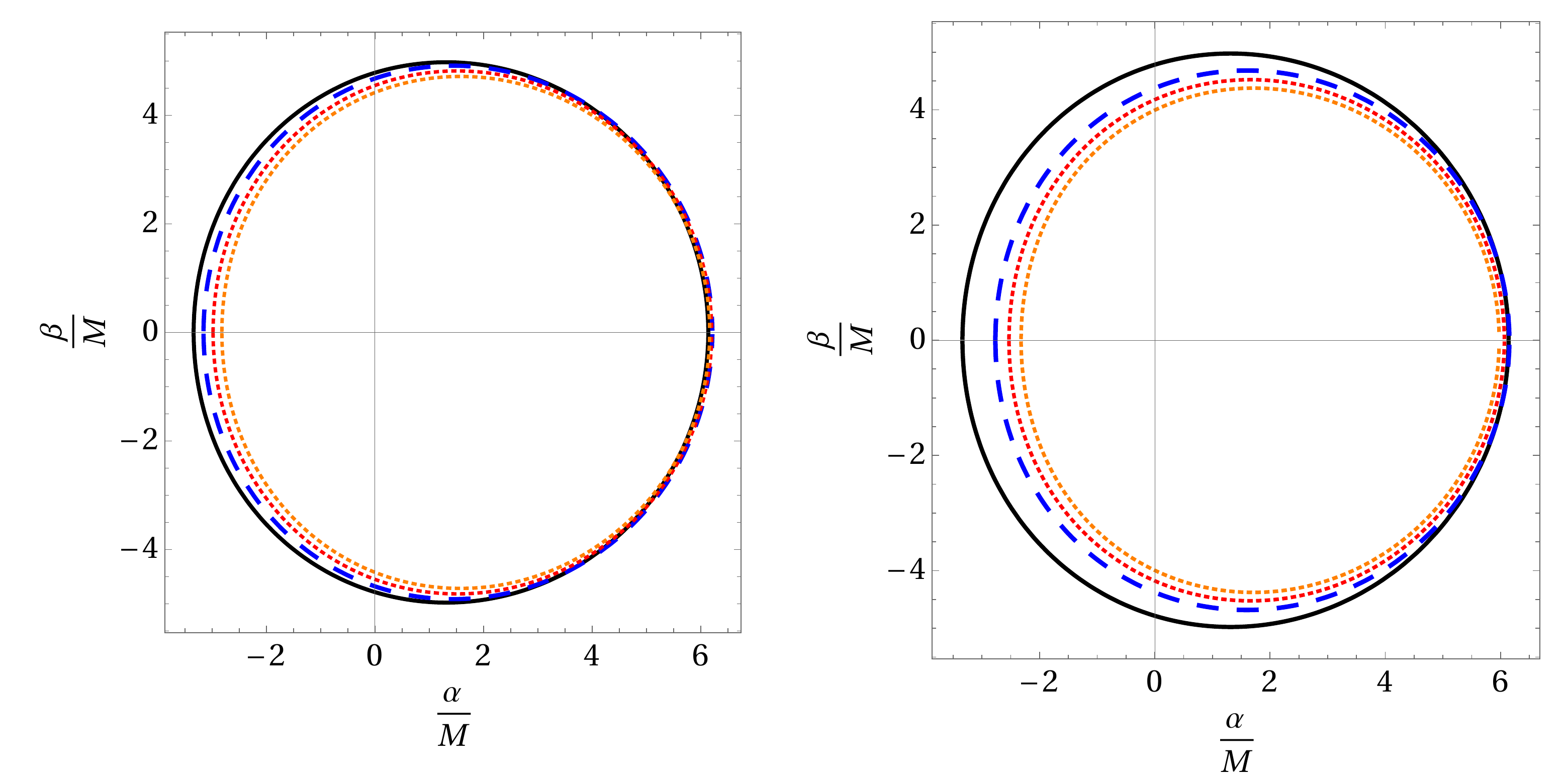}
\caption{\label{fig22} 
Silhouette of the shadow cast by the rotating SD BH for $a=0.75$ with $\Lambda=-0.01$, $M=1$, $G_{0}=1$
and $c_{4}=10$. In the left panel $\epsilon=0.00$ (black solid line), $\epsilon=0.03$ (blue dashed line) and
$\epsilon=0.06$ (red dotted line) and $\epsilon=0.09$ (orange dotted line). In the right panel
$\epsilon=0.00$ (black solid line), $\epsilon=0.10$ (blue dashed line) and
$\epsilon=0.15$ (red dotted line) and $\epsilon=0.20$ (orange dotted line)
 }
\end{figure*}

The horizons of the solution correspond to the real roots of $\Delta$. However,
given the nature of this function, numerical computations are required. In figure \ref{fig1} we show the behaviour of $\Delta$ for 
 fixed values of $\{M,G_{0},a,\Lambda\}$ parametrized by the running parameter,$\epsilon$.
It is worth noticing that the size of the BH decreases as the running parameter takes greater values.
Causality violation and closed time--like 
curves are possible if $g_{\phi\phi}>0$,  namely
\begin{eqnarray}
-\frac{\Sigma  \sin ^2(\theta )}{\rho ^2}>0,
\end{eqnarray}
from where, given that $\sin^{2}(\theta)/\rho ^2$ is positive, the sign of $\Sigma$ plays a crucial role in the analysis. What is more, the condition to avoid causality violation and closed time--like curves is to impose $\Sigma>0$. In figure \ref{figvc}, it is shown that, in contrast to Kerr solution, this requirement is fulfilled for certain values of the parameters.
Note that for the particular values of $\{a,M,\Lambda,\epsilon\}$ setted in \ref{figvc}, $\Sigma$ is a monotonously increasing function of $r$ but is shifted downwards for increasing values of the running parameter. Even more, it can be demonstrated that there exists an upper bound on $\epsilon$ above which the emergence of a local maximum and, as a consequence,  the apparition of negatives values of $\Sigma$ is unavoidable. However, given the functional form of $\Sigma$, the bound can only be estimated by numerical calculation. In the case shown in figure \ref{figvc} the upper bound is around $\epsilon\approx 0.89$.
\\
The BH shadow corresponds to the parametric function of the celestial coordinates. In figure \ref{fig2} 
the shadow of the rotating BH is shown for different values of the parameters.
Note that for small values of the running parameter, $\epsilon<0.1$,  the shadow undergoes a shift to the left without an important change in its size (Fig. \ref{fig2}, left panel). However, for greater values of epsilon (Fig. \ref{fig22}, right panel), the effect is on the size and the shape of the shadow. A similar effect occurs when considering another set of parameters. In particular, for a higher value of the spin parameter, $a$,  the effect on the shadow is shown in Fig. \ref{fig22}.
Note that in this case for $\epsilon<0.1$ the shift of the shadow coincides with the obtained in the previous case, but the effect is remarkable. The same occurs for $\epsilon>0.1$, namely, the deformation of the shadow is noteworthy.\\

Before concluding this section
	we would like to discuss some issues we find when the NJA algorithm is implemented regarding to either the final solution solve the Einstein equations or not as was pointed out at the end of section \ref{NJ}. To this end we shall follow the statements in references \cite{rodrigues2017} and \cite{bobirmod}. In these works, the authors claim that the introduction of the rotation parameter by the NJA algorithm must change the form of the nonlinear electrodynamics Lagrangain and its total derivative respect to the Maxwell scalar, $\mathcal{L}(F)$ and $d\mathcal{L}(F)/dF$ respectively. However, the only way in which we can obtain such a modification is using the rotating solution of the Einstein field equations as a background to obtain the the form for $\mathcal{L}(F)$ and $d\mathcal{L}(F)/dF$. Nevertheless, this strategy leads to five independent equations that can not be consistently solved and in this sense, the solution obtained by the NJA must be considered as an approximation. Regardingly, in this work, the context of scale--dependent gravity, the Einstein field equations, in vacuum, are given by Eq. (\ref{gmn}).
 Note that, Eq. (\ref{gmn}), can be written as
	\begin{eqnarray}\label{Gmn}
	G_{\mu\nu}=\tilde{T}_{\mu\nu}
	\end{eqnarray}
	with
	\begin{eqnarray}
	\tilde{T}_{\mu\nu}=-\Delta t_{\mu\nu}-g_{\mu\nu}\Lambda.
	\end{eqnarray}	
	Now, (as stated in ref. \cite{bobirmod} in the context of nonlinear electrodynamics) the introduction of 
	the rotation parameter by the Newman--Janis algorithm must change the form of the fields $G$ and $\Lambda$ given by Eqs. (\ref{G}) and (\ref{L}), respectively.  However, it is not possible to apply the Newman--Janis algorithm directly in $\tilde{T}_{\mu\nu}$. The only way to find $G(r,\theta)$ and $\Lambda(r,\theta)$ of the rotating black hole, is to solve Einstein field equations $G_{\mu\nu}=\tilde{T}_{\mu\nu}$ with respect to $G(r,\theta)$ and $\Lambda(r,\theta)$ in the background defined by the rotating metric (\ref{blsm}) Now, an explcit calculation reveals that Eq. (\ref{Gmn}) leads to five independent equations involving derivatives of of $G(r,\theta)$ and constrained to
	\begin{eqnarray}
	&&G(r,\theta ) \left(2 \partial^{2}_{r\theta}G(r,\theta )-\frac{\partial_{r}G(r,\theta ) \partial_{\theta}g_{rr}}{g_{rr}}\right)\nonumber\\
	&&-\partial_{\theta}G(r,\theta ) \left(4 \partial_{r}G(r,\theta )+\frac{G(r,\theta ) \partial_{r}g_{\theta\theta}}{g_{\theta\theta}}\right)=0,
	\end{eqnarray}
	which arise after imposing $\tilde{T}_{r\theta}=0$. However, up to now, no analytical solution to this system is known.	
	In this sense, we conclude that it is not possible to solve consistently all the equations to obtain $G(r,\theta)$ and $\Lambda(r,\theta)$ and, as a consequence, it is not possible to demonstrate if the solution corresponds to a solution of the Einstein field equations. However we would like to point out that, 
	independent of whether this metric could be a solution or not, it is interesting to see the lensing of light in this geometry as a case study as we did in this work.

\section{Conclusions}\label{remarks}

In this work, we have reviewed some aspects on the Newman--Janis algorithm without complexification that allows us to construct generic four--dimensional rotating black holes. In particular, we have constructed a rotating scale--dependent black hole, and we have studied the main aspects of the construction of its unstable null orbits. These are new results since to the best of our knowledge this has not been considered before. Moreover, some physical properties, such as the position of the horizons, the static limit, and causality issues, have been investigated. 

\smallskip

One of the main results obtained here is that the running parameter induces significant variations in the behavior of the rotating solutions in comparison with its classical counterpart. To be more precise, we found that the size of the rotating scale--dependent black hole decreases as the running parameter takes greater values. Moreover, we demonstrated that, in contrast to the Kerr solution, the causality issues can be circumvented for certain values of the free parameters of the solution. What is more, for fixed $\{a,M,\Lambda\}$ we can obtain a bound on the running parameter to ensure the positivity of $\Sigma$. We also obtained that the black hole shadow is affected by the running parameter. Indeed, the shadow undergoes a deformation and its size is reduced as the running parameter increases.

\smallskip

We would like to emphasize that the Newman--Janis algorithm implemeted here is free of pathologies (in the sense described in Ref.\cite{rodrigues2017}) that could appear when a matter Lagrangian is considered into the action. Indeed, as we have commented after Eq. (\ref{Gg}), there is not any matter field  in the system under study. Even more, the equations of motion we solved correspond to the scale-dependent Einstein field equations that arise as a consequence of variations of the scale—dependent action respect to the metric field. To be more precise, the Lagrangian density appearing in Eq. (\ref{actionG}) vanishes and the only contribution to the effective energy momentum tensor is given by the so--called non-matter energy momentum tensor given by Eq. (\ref{Gg}).

\smallskip

Finally, it would be interesting to obtain the scale--dependent Newton's constant and cosmological constant for the rotating solution, namely
$G(r,\theta)$ and $\Lambda(r,\theta)$ respectively, after solving the scale--dependent Einstein's equation 
	\begin{equation}
	G_{\mu \nu } + \Lambda(r,\theta) g_{\mu \nu} = 8\pi G(r,\theta) T_{\mu \nu}- \Delta t_{\mu \nu},
	\end{equation}
considering (\ref{blsm}) as the line element of the solution. However, as it was not the main goal of this work, we hope to consider this and other issues in future work.

\section*{Acknowlegements}

The author \'A.~R. acknowledges DI-VRIEA for financial support through Proyecto Postdoctorado 2019 VRIEA-PUCV. The author G.~P. thanks the Funda\c c\~ao para a Ci\^encia e Tecnologia (FCT), Portugal, for the financial support to the Center for Astrophysics and Gravitation-CENTRA, Instituto Superior T{\'e}cnico,  Universidade de Lisboa, through the Grant No. UID/FIS/00099/2013.


\end{document}